\documentclass[twocolumn,aps,prb,superscriptaddress,notitlepage,longbibliography]{revtex4}
\usepackage[colorlinks=true,urlcolor=blue,citecolor=blue,linkcolor=blue]{hyperref}
\usepackage{graphicx}  
\usepackage{dcolumn}   
\usepackage{bm}        
\usepackage{amssymb}
\usepackage{amsfonts}
\usepackage{verbatim}
\usepackage{epstopdf}
\usepackage{amsmath}
\usepackage{braket}
\usepackage{hyperref}

\begin{document}

\title{ Disorder-induced phase transitions in higher-order nodal line semimetals}
\author{Yue-Ran Ding}
\affiliation{Institute for Advanced Study and School of Physical Science and Technology, Soochow University, Suzhou 215006, China}
\author{Dong-Hui Xu}\thanks{donghuixu@cqu.edu.cn}
\affiliation{Department of Physics and Chongqing Key Laboratory for Strongly Coupled Physics, Chongqing University, Chongqing 400044, China}
\affiliation{Center of Quantum Materials and Devices, Chongqing University, Chongqing 400044, China}
\author{Chui-Zhen Chen}\thanks{czchen@suda.edu.cn}
\affiliation{Institute for Advanced Study and School of Physical Science and Technology, Soochow University, Suzhou 215006, China}

\begin{abstract}
Higher-order nodal line semimetals represent a recently proposed topological semimetal class that harbors bulk nodal lines and features gapless hinge Fermi arc excitations, governed by the bulk-hinge correspondence. In this study, we investigate the disorder effect on a higher-order nodal line semimetal and the consequent phase transitions. Within the pristine higher-order nodal line semimetal model, we unveil three distinct phases: higher-order nodal line semimetal, conventional nodal line semimetal, and normal insulator. The higher-order nodal line semimetal is characterized by one-dimensional hinge Fermi arc states connecting a pair of nodal rings, contrasting with conventional nodal line semimetals that exhibit two-dimensional drumhead surface states. We demonstrate that disorder can trigger multiple phase transitions within this system. Significantly, intermediate disorder can induce higher-order topology in an initial conventional nodal line semimetal or even an initial normal insulator. Further increase in disorder drives the system through a diffusive metallic phase before ultimately reaching the Anderson insulator regime. Employing a combination of finite-size scaling analysis and an effective medium theory, we construct a comprehensive phase diagram, elucidating the intricate interplay between disorder and topology.

\end{abstract}

\maketitle

\section{Introduction}
In recent years, the field of topological phases of matter has emerged as a highly dynamic area,  driven by the discoveries of gapped topological insulators~\cite{TIs-RMP,TIs-TSs-RMP} and gapless topological semimetals~\cite{WeylDirac-RMP}. Based on the geometric structure of their band touching points, topological semimetals can be categorized as two main types: point-like and line-like~\cite{BAAPRB2011}. Weyl semimetals~\cite{wanprb2011wsm,murakami2007phase} are a type of highly celebrated point-like topological semimetal. Nodal line semimetals facilitate low-energy excitations along one-dimensional nodal lines arising from the intersection of the conduction and valence bands~\cite{TDPRB2016,LRPRB2016,BGNC2016,SLMNC2016,HMMPRB2017,WXBPRB2017,LZHPRX2018,GMPRL2020}. This characteristic sets them apart from semimetals characterized by point-like configurations. Meanwhile, two-dimensional (2D) `drumhead' surface states exist on surfaces parallel to the nodal line plane, which are regarded as a key manifestation of the topological nature of NLSMs \cite{BAAPRB2011,WHMPRB2015,CYHPRB2016,ZDWPRA2016}.
So far, plenty of NLSM materials have been experimentally observed~\cite{XLSAPLM2015,XCQPRM2017,LRnpjQM2018,SCPRM2020,BGNC2016,SLMNC2016,HJPRL2016,HJPRB2017,NTPRB2019,OYJotPSoJ2016,LAPRB2019,QZYPRB2019}.

More recently, higher-order topological phases have been proposed as a generalization of the concept of topological phases of matter \cite{WABS2017,KBPRB2019,CRPRL2020,DYRPRB2020}. A $d$-dimensional $n$th-order topological insulator possesses gapless boundary modes on the $(d-n)$-dimensional boundaries with $1 < n \leq d$, unlike a conventional first-order topological phase for which $n = 1$.
Subsequently,  the notion of higher-order topology was  extended to gapless semimetallic systems \cite{EPRL2018, LMPRB2018, RPRR2019,WBJNC2020,WZMPRB2023,EPRB2018,WHXPRL2020,GPRL2020,GSAAPRL2020,WQNM2021,RWBnpj2022}. Interestingly, the higher-order nodal line semimetal (HONLSM) hosts nodal rings in the bulk and 1D gapless hinge Fermi arc states terminated on the projection of pairs of nodal rings~\cite{AJPRL2018,WZJPRL2019,WKPRL2020,LEnpj2020,ZYXPRL2021,DXLPRB2022}, which is in contrast to the first-order NLSMs only supporting 2D `drumhead' surface states.

On the other side, disorder has a major impact upon the transport properties of low-dimensional electronic systems, as highlighted by P. W. Anderson in 1958 \cite{AndersonPR1958}. The disorder-induced metal-insulator transitions, known as the Anderson transitions, have been extensively studied in various materials. The Anderson transition has been also studied in topological semimetals, including WSMs \cite{CCZPRL2015}, double Weyl semimetals \cite{ZJYPRB2022}, and NLSMs \cite{LXLPRB2020, ZYXarXiv2022}, where the interplay between band topology and disorder gives rise to more exotic localization phenomena. For instance, in a disordered Weyl semimetal, Weyl nodes can annihilate pairwise when they approach each other in momentum space or by strong intervalley scattering at finite disorder strength \cite{CCZPRL2015}.
It is found that an infinitesimally small disorder drives the nodal line Dirac semimetal in the clean limit to the metal~\cite{LXLPRB2020}.
More recently, the study of the disorder-induced phase transitions of higher-order topological phases also has drawn intensive attention
\cite{AHPRB2019,SZXCPB2019,LCAPRL2020,AAPRR2020,HYSPRB2021,YBYPRB2021,ZWXPRL2021,WCPRR2020,WCPRB2021}.
So far, most of the studies have focused on high-order topological insulators, and investigations into the disorder effects of higher-order topological semimetals are still scarce \cite{SPRR2020,ZZQPRB2021A}.
Notably, the role of disorder in HONLSMs hasn't been thoroughly investigated yet and needs to be further explored.

In this paper, we study the topological characteristics and disorder-driven phase transitions in a HONLSM model. The pure HONLSM model is a four-band Dirac Hamiltonian on cubic lattice, which exhibits three distinct phases including HONLSM, NLSM, and normal insulator (NI) as the Dirac mass varies. We determine their phase boundaries by computing the energy spectrum,  topological invariants, and typical state distributions. The hallmark feature of the HONLSM phase is the existence of zero-energy hinge Fermi arc~\cite{WKPRL2020} states residing on the different hinges, manifesting the higher-order bulk-boundary correspondence. Introducing disorder that preserves both $PT$ and chiral symmetries, we map out a global phase diagram using the transfer matrix method and the self-consistent Born approximation (SCBA). Within the framework of SCBA, the disorder normalizes and reduces the Dirac mass, giving rise to phase transitions such as NLSM-HONLSM or NI-NLSM-HONLSM transitions. In the presence of strong disorder, a diffusive metal (DM) emerges before the Anderson insulator (AI) phase is realized.

\section{ Model Hamiltonian}\label{II}
To start with, we consider a spinless four-band model on a 3D cubic lattice, and the tight-binding Hamiltonian in momentum space reads\cite{WKPRL2020,DXLPRB2022}
\begin{eqnarray}
H(\textbf{k})=M_k\gamma_3+t\sin k_x\gamma_1+t\sin k_y\gamma_2+H_{\mathrm{mass}},
\label{eq1}
\end{eqnarray}
where $M_k=M-t(\cos k_x+\cos k_y+ \cos k_z)$, $\gamma_1=\rho_0\otimes\tau_3$, $\gamma_2=\rho_2\otimes\tau_2$, $\gamma_3=\rho_0\otimes\tau_1$ and $\gamma_4=\rho_1\otimes\tau_2$ are the $4\times4$ Dirac matrices. $\rho_i$ and $\tau_i$($i=1,2,3$) are the Pauli matrices, and $\rho_0$ is the $2\times2$ identity matrix. $\rho_i$ and $\tau_i$ act on layer and sublattice spaces. $M$ describes the Dirac mass and $t$ is the hopping amplitude. The last term $H_{\mathrm{mass}}=i\Delta(\gamma_1\gamma_4+\gamma_2\gamma_4)$ represents an additional mass which generates nodal rings and higher-order topology. We set $t=1$ throughout the paper for simplicity.

The Hamiltonian in Eq.~(\ref{eq1}) respects inversion symmetry with inversion operator $P=\gamma_3$, and time-reversal symmetry with operator $T=\gamma_3 \mathcal{K}$, where $\mathcal{K}$ denotes complex conjugate.
In addition, the Hamiltonian also has chiral symmetry $C=\gamma_4$ under which the Bloch Hamiltonian obeys $CHC^{-1}=-H$ and mirror symmetry $M_{x\bar{y}}=\gamma_1\gamma_4+\gamma_2\gamma_4$ on the mirror plane $k_y=-k_x$ with $M_{x\bar{y}}H(\textbf{k})M_{x\bar{y}}^{-1}=H(-k_y,-k_x,k_z)$.

In the absence of $H_{\mathrm{mass}}$, the Hamiltonian describes a 3D DSM with energy spectrum give by $E_k = \pm\sqrt{M_k^2 + t^2\sin^2k_x+t^2\sin^2k_y}$, and the locations of Dirac nodes depend on $M$. When $M\in[1,3]$, there is one pair of the Dirac nodes located at $[0,0,\pm\arccos(M-2)]$.
For $M\in[-1,1]$, two pairs of the Dirac nodes reside at $[0,\pi,\pm\arccos M]$ and $[\pi,0,\pm\arccos M]$. Similarly, they are located at $[\pi,\pi,\pm\arccos(M+2)]$ in the region $M\in[-3,-1]$.
By turning on $H_{\mathrm{mass}}$, the two Dirac nodes evolve into two nodal rings on the mirror plane $k_x=-k_y$, signaling the transition from a DSM to an NLSM.
Note that the sizes of nodal rings are related to $\Delta$. In the following calculations, we take $\Delta=0.2$.

\section{Topological features and electronic structures of the pure system}\label{III}
\begin{figure}[tbh]
	\centering
	\includegraphics[width=3.4in]{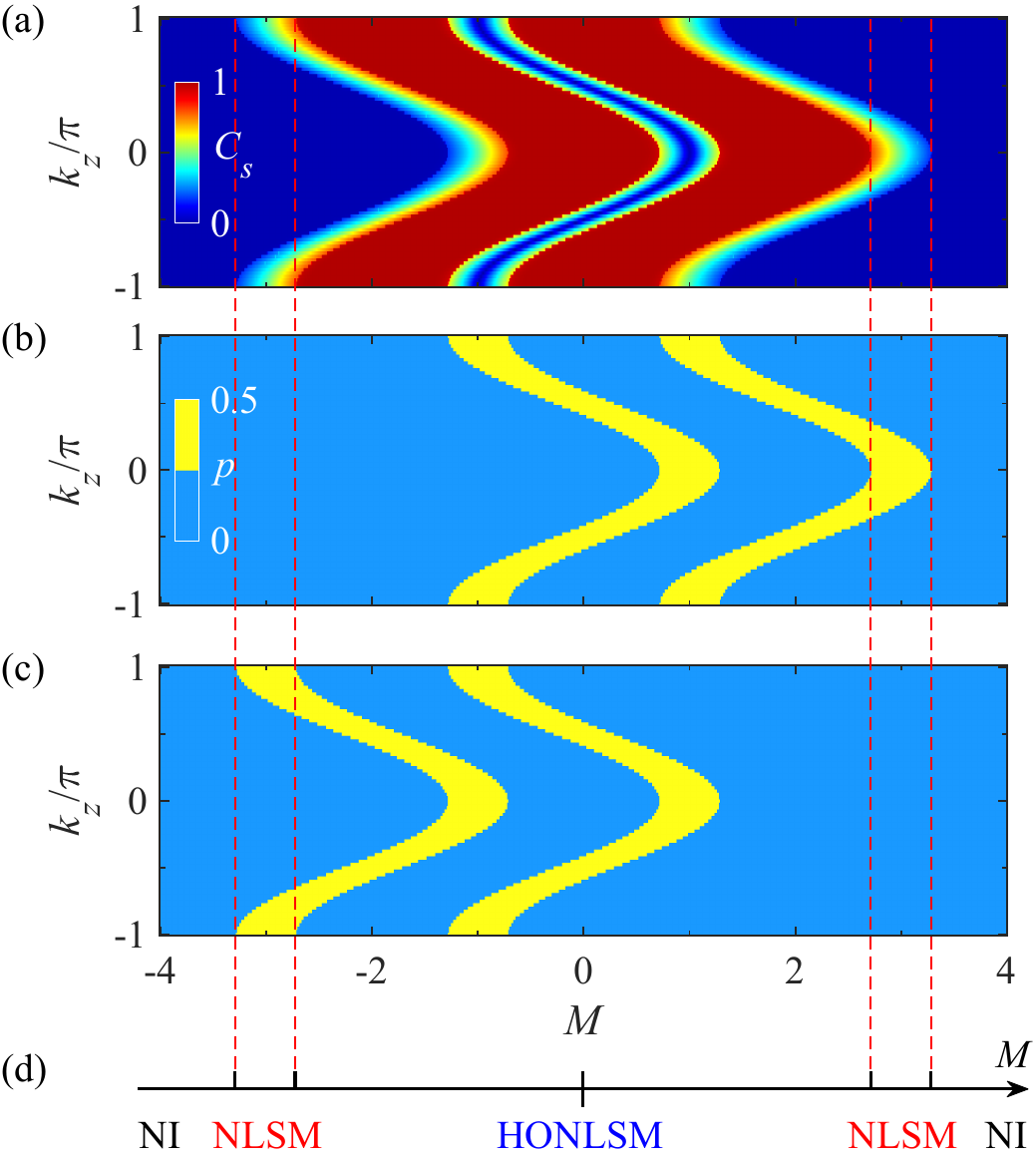}
	\caption{(Color online). (a) The spin Chern number $C_s$ for various Dirac mass $M$ and momentum $k_z$ with the periodic boundary condition in $x$ and $y$ directions, and $N_x=N_y=30$. Bulk polarization $p$ versus $M$ and $k_z$ for (b) $k_y=0$ and (c) $k_y=\pi$, respectively. (d) Phase diagram of the pure Dirac model. For different $M$, there are HONLSM, NLSM and normal insulator (NI) phases. The region between two red dashed lines is the NLSM phase. Note that the range of the NLSM region is proportional to $\Delta$. The parameter is set to be $\Delta=0.2$ throughout the paper.
		\label{fig1} }
\end{figure}

To capture the topological characteristics of the clean Dirac model, we calculate the spin Chern number and the bulk polarlization. As mentioned in Sec. \ref{II}, the DSM is the parent phase of the NLSM and HONLSM. If we treat $k_z$ as a model parameter, then the DSM model ($H_{\mathrm{mass}}=0$) reduces to a 2D topological insulator model whose topology can be characterized by the spin Chern number. When $H_{\mathrm{mass}}$ is turned on, we can still have a quantized spin Chern numnber except the $k_z$ plane that has the projection of nodal rings.
The spin Chern number of 2D system with sample size $N_x\times N_y$ and certain momentum $k_z$ is defined as $C_s=|C_+-C_-|/2$, with the Chern number $C_{\pm}$ of two different subsystems based on the non-commutative Kubo formula can be expressed as $C_{\pm} = 2\pi i\mathrm{Tr}[Q_{\pm}[\partial_{k_x} Q_{\pm},\partial_{k_y}Q_{\pm}]]$\cite{ProdanPRB2009,HYSPRB2021}. $Q_{\pm}$ is spectral projector onto the positive/negative eigenvalue of $Q(\rho_2\otimes\tau_0)Q$, where $Q$ is the projector onto the occupied states of $H$ with the periodic boundary condition (PBC) in $x$ and $y$ directions.

Meanwhile, the bulk polarization of the system can be determined by
the Wilson loop method \cite{WFPRL1984,AAPRB2014,WABS2017,BWAPRB2017,FSPRB2018,LXWPRL2019}.
For given $k_y$ and $k_z$, the Wilson loop operator alone the path in the $k_x$-direction defined as
\begin{eqnarray}
W_{x,\textbf{k}}=F_{x,\textbf{k}+(N_x-1)\Delta_{k_x}}...F_{x,\textbf{k}+\Delta_{k_x}}F_{x,\textbf{k}}, \nonumber
\end{eqnarray}
where  $[F_{x,\textbf{k}}]^{mn}=\langle u_{\textbf{k}+\Delta_{k_x}}^m|u_{\textbf{k}}^n\rangle$, with occupied Bloch wave functions $|u_{k_x}^n\rangle$ for $n=1,2,...,N_{\mathrm{occ}}$, and the number of occupied bands $N_{\mathrm{occ}}$.
Here $\textbf{k}=(k_x,k_y)$ denotes the initial point of the Wilson loop operator, and $\Delta_{k_x}=2\pi/N_x$ represents the step with $N_x$ unit cells in $x$-direction.
The eigenvalue equation of the Wilson loop operator reads $W_{x,\textbf{k}}|\nu_{x,\textbf{k}}^j\rangle=e^{i2\pi\nu_x^j(k_y)}|\nu_{x,\textbf{k}}^j\rangle$, where $j=1,2,...,N_{\mathrm{occ}}$.
The polarization as a function of $k_y$ is given by the sum of the phases $\nu_x^j(k_y)$:
\begin{eqnarray}
p(k_y)=\sum_j\nu_x^j(k_y)\mod1=-\frac{1}{2\pi}\log\det[W_{x,\textbf{k}}].\nonumber
\end{eqnarray}
The polarization  between two nodes is quantized to $1/2$ \cite{LZRPRB2021}.

We plot the spin Chern number $C_s$ and bulk polarization $p$ versus $M$ and $k_z$ in Figs.~\ref{fig1}(a)-\ref{fig1}(c), which gives rise the phase diagram of the clean system shown in Fig.~\ref{fig1}(d). Figures~\ref{fig1}(b) and \ref{fig1}(c) depict $p$ at $k_y=0$ and $k_y=\pi$, respectively. The unquantized regions of $C_s$ are consistent with the regions where $p$ is quantized to $1/2$, that is, the regions lie nodal rings, wherein exist the drumhead surface states. The situations of $M<0$ are analogous to those of $M>0$, thus in the subsequent discussions, we will mainly focus on the cases $M\geqslant0$. In the region between two red dashed lines of Fig.~\ref{fig1}, there is only one bulk nodal ring, corresponding to the conventional NLSM phase.

\begin{figure}[tbh]
	\centering
	\includegraphics[width=3.4in]{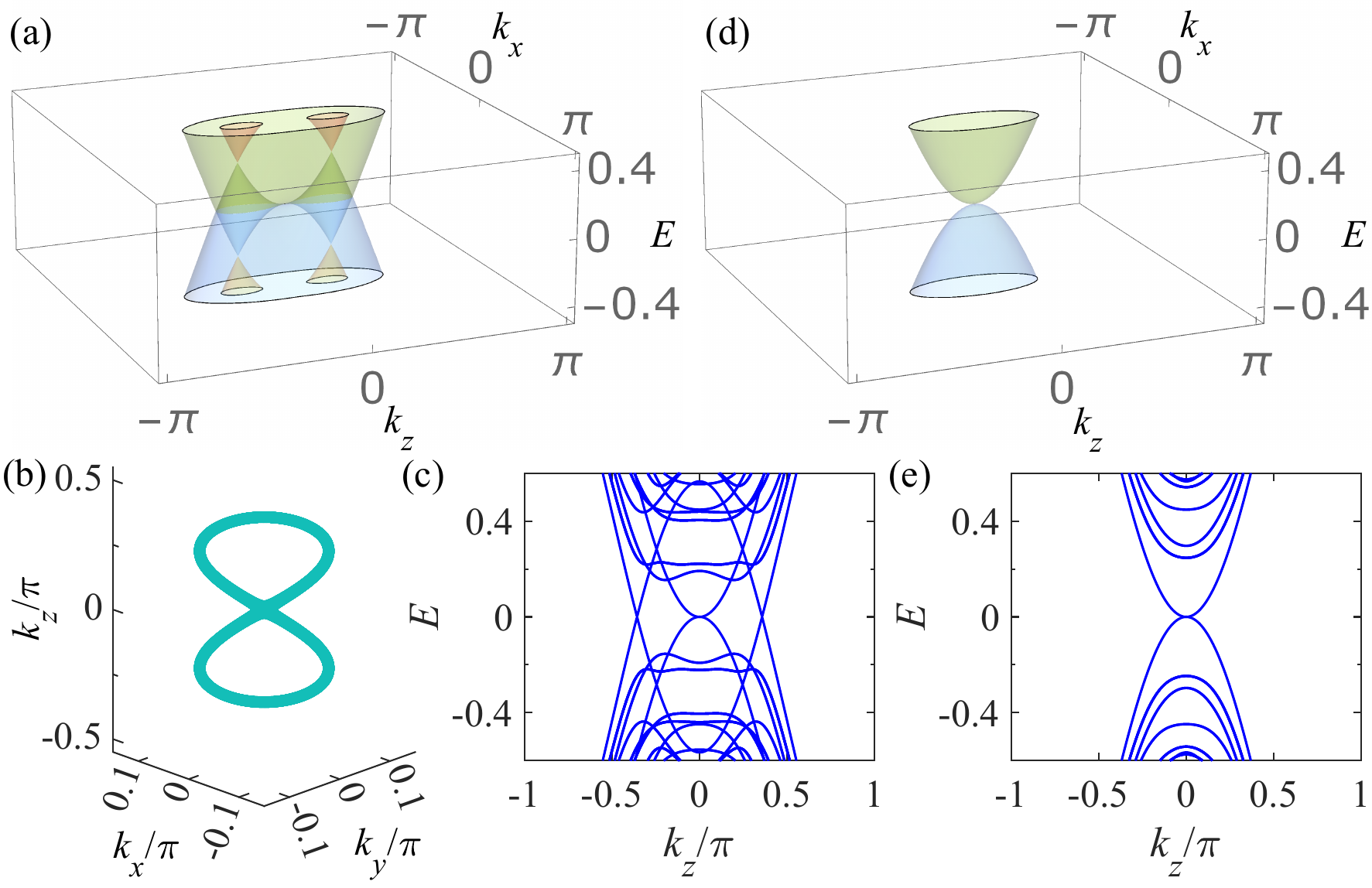}
	\caption{(Color online). Electronic structures of phase boundaries of the NLSM phase. $M=M_-=3-\sqrt2\Delta$ for (a)-(c). In (d)-(e) $M=M_+=3+\sqrt2\Delta$.
(a) and (d) are the bulk energy spectra on the $k_z-k_x$ plane with $k_y=-k_x$. (c) Two connected nodal rings in the 3D Brillouin zone.
(c) and (e) Bulk energy bands with size $N_x=N_y=20$.
		\label{fig2} }
\end{figure}
Next, we proceed to identify the phase boundaries through a detailed exploration of the band structure's evolution. The critical point of phase transition is linked to the zero-energy spectrum of the Hamiltonian  $H$ in Eq.~\ref{eq1}, specifically represented by the solution to the equation:
$[M_\textbf{k}^2+(\sin^2 k_x-\Delta^2)+(\sin^2 k_y-\Delta^2)]^2+4\Delta^2(\sin k_x+\sin k_y)^2=0$.
This implies that $\sin k_y=-\sin k_x$, and leads further to the following equation:
\begin{eqnarray}
(M-2\cos k_x-\cos k_z)^2+2(\sin^2 k_x-\Delta^2)=0.
\label{eq2}
\end{eqnarray}
For positive $M$, the gap at $\textbf{k}=(0, 0, 0)$ closes when $M=M_-=3-\sqrt2\Delta$ [see Figs.~\ref{fig2}(a)-\ref{fig2}(c)] and $M=M_+=3+\sqrt2\Delta$ [see Figs.~\ref{fig2}(d) and \ref{fig2}(e)].
In Figs.~\ref{fig2}(a) and \ref{fig2}(d), we plot the bulk energy spectra as a function of $k_z$ and $k_x$ with $k_y =-k_x$ for $M=M_\pm$.
Meanwhile, the bulk energy bands versus $k_z$ with PBC in both $x$ and $y$ directions at $M=M_\pm$ are shown in Figs.~\ref{fig2}(c) and \ref{fig2}(e).
As depicted in Fig.~\ref{fig2}(c), when $M=M_-$, two nodal rings in 3D Brillouin zone connect with each other [see also Figs.~\ref{fig2}(a)-\ref{fig2}(b)].
As for $M=M_+$, the nodal ring shrinks into a node point [see Figs.~\ref{fig2}(d) and \ref{fig2}(e)].
As a result, one obtain that the parameter range of the NLSM phase with a single nodal ring is $M_+-M_-=2\sqrt2\Delta$.

\begin{figure}[tbh]
	\centering
	\includegraphics[width=3.4in]{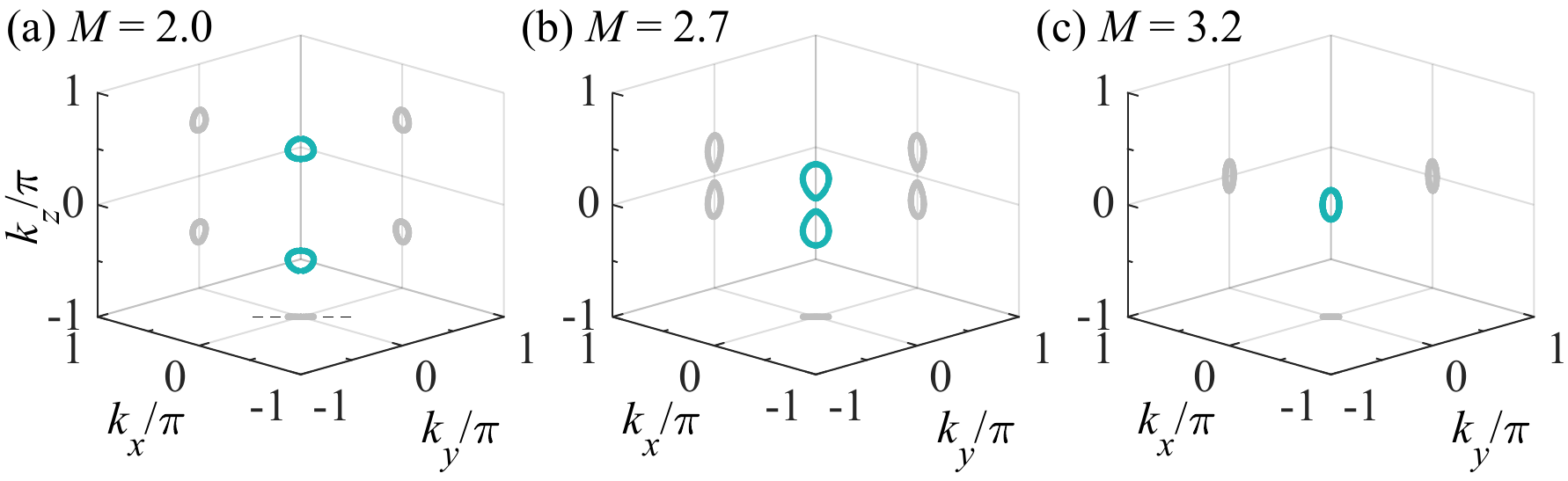}
	\caption{(Color online). Nodal rings in the 3D Brillouin zone.
The gray rings are the projections of the nodal rings on the $k_x-k_y$, $k_y-k_z$, and $k_x-k_z$ planes, respectively. The dashed line in (a) represents $k_y=-k_x$ direction.
We take $M=2.0$ for (a), $M=2.7$ for (b), and $M=3.2$ in (c). (a) and (b) are the HONLSM with two nodal rings, while (c) is the NLSM phase with only one nodal ring.
		\label{fig3} }
\end{figure}
To uncover the evolution of the nodal rings with the change in Dirac mass $M$,
we plot the nodal rings in the 3D Brillouin zone for $M=2.0, 2.7$ and $3.2$ in Figs.~\ref{fig3}(a)-\ref{fig3}(c), respectively. 
As $M$ increases, two nodal rings of the HONLSM come closer together [Figs.~\ref{fig3}(a) and \ref{fig3}(b)], shortening the hinge Fermi arcs between them and increasing the localization length of  the hinge Fermi arc states. 
When $M = M_-$, two nodal rings merge into one [see Fig.~\ref{fig2}(b)]. Consequently, the hinge Fermi arcs disappear, signifying that $M_-$ is the critical point for the HONLSM-NLSM transition.
The nodal ring of the NLSM [Fig.~\ref{fig3}(c)] shrinks with the increase of $M$, and it annihilates at $M=M_+$ [Fig.~\ref{fig2}(d)], which represents the NLSM-NI transition point.
More discussions of the HONLSM phase are shown in Appendix~\ref{AB}.

\begin{figure}[tbh]
	\centering
	\includegraphics[width=3.4in]{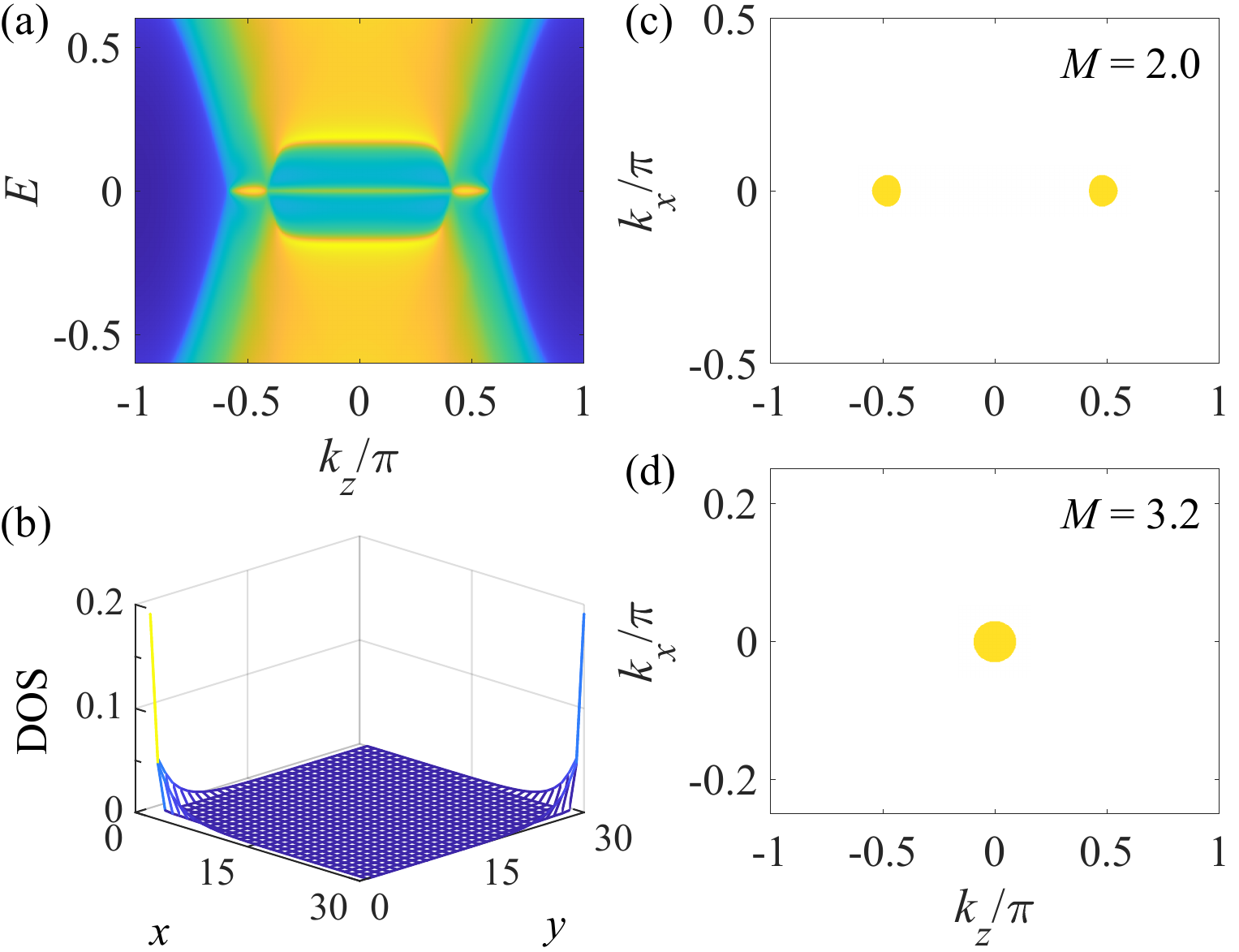}
	\caption{(Color online). Electronic structure of (a)-(c) HONLSM phase with $M=2.0$ and (d) NLSM phase with $M=3.2$.
(a) DOS versus momentum $k_z$ and energy $E$ for HONLSM system with open boundary conditions (OBC) in $x$ and $y$ directions and size $N_y=200$.
(b) DOS in the $x$-$y$ plane of real space of the hinge Fermi arc states at $k_z=0$ with sample size $N_x\times N_y=30\times 30$.
The spectral density at (c) $M=2.0$ and (d) $M=3.2$ in the surface Brillouin zone defined on the $k_x$-$k_z$ plane with $E=0$. The bright yellow region represent the drumhead surface states bounded by the nodal rings.
The dark yellow line between two nodal rings (bright yellow) at $E=0$ in (a) represents the hinge Fermi arc.
The density in (a), (c) and (d) are calculated by the surface Green's function method.
		\label{fig4} }
\end{figure}

Next, we compare the topological characteristics of the HONLSM [Figs.~\ref{fig4}(a)-\ref{fig4}(c)] and the NLSM [Fig.~\ref{fig4}(d)] by the calculation of boundary states.
Figures~\ref{fig4}(a) shows the spectral function as a function of $k_z$ and $E$ for the HONLSM with OBC along $x$ and $y$ directions.
As shown in Fig.~\ref{fig4}(a), a pair of hinge Fermi arc states with zero-energy flat bands [see the dark yellow line at $E=0$] connect the two nodal rings. As illustrated in Fig.~\ref{fig4}(b), we demonstrate the hinge Fermi arc states ($k_z-0$) of a 2D $N_x\times N_y=30\times 30$ sample  and $k_z=0$ are located on two mirror-symmetric off-diagonal hinges. 
Additionally, We plot spectral function of $E=0$ in the surface Brillouin zone defined on the $k_x$-$k_z$ plane as shown in Figs.~\ref{fig4}(c) and \ref{fig4}(d), respectively. We can see that there exist surface drumhead states within the projection of both two bulk nodal rings for HONLSM [Fig.~\ref{fig4}(c)], corresponding to the bright yellow lines in Fig.~\ref{fig4}(a) at $E=0$. In comparison, we also show the surface drumhead states within the projection of single nodal ring for the conventional NLSM [see Fig.~\ref{fig4}(d)]. 
The spectral functions in Figs.~\ref{fig4}(a), \ref{fig4}(c) and \ref{fig4}(d) are determined by the surface Green's function method with an infinitesimal imaginary part of Fermi energy set to be $\eta=10^{-2}$, while the DOS in Fig.~\ref{fig4}(b) is calculated by the eigenstate wavefunctions of the Hamiltonian in Eq. (\ref{eq1}). 

\section{Phase diagram of Disordered Dirac System}
\begin{figure}[tbh]
	\centering
	\includegraphics[width=3.4in]{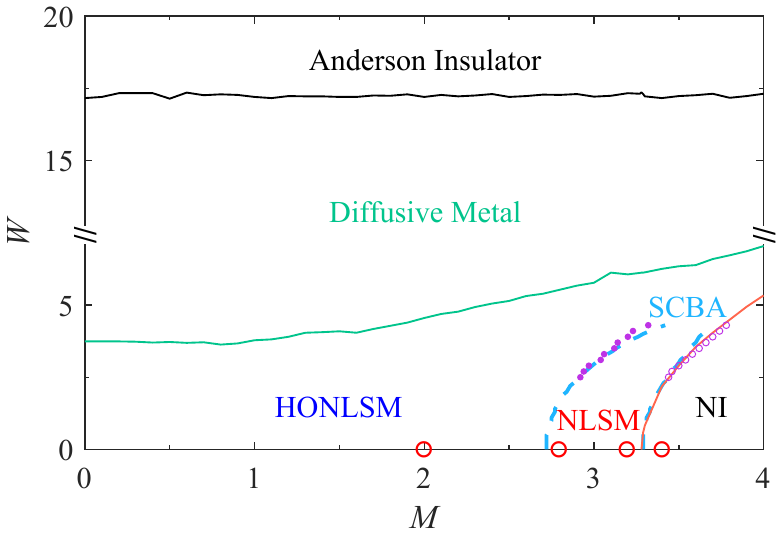}
	\caption{(Color online). Phase diagram of disordered Dirac system on the $W-M$ plane. The solid lines are determined by localization length. The dashed lines are the phase boundaries obtained by utilizing SCBA. The violet dots and circles are approximately determined by the critical points of $\Lambda-M$ curve [Fig.~\ref{fig7}]. The red circles are four typical points.
		\label{fig5} }
\end{figure}
In this section, we study the disorder effect on the Dirac system. The on-site Anderson disorder that preserves the $PT$ and chiral symmetries is described as $H^d = \sum_j\varepsilon_j \rho_3 \otimes \sigma_0$, where $\varepsilon_j$ is uniformly distributed within $[-W/2, W/2]$ with a disorder strength of $W$.
\begin{figure}[tbh]
	\centering
	\includegraphics[width=3.4in]{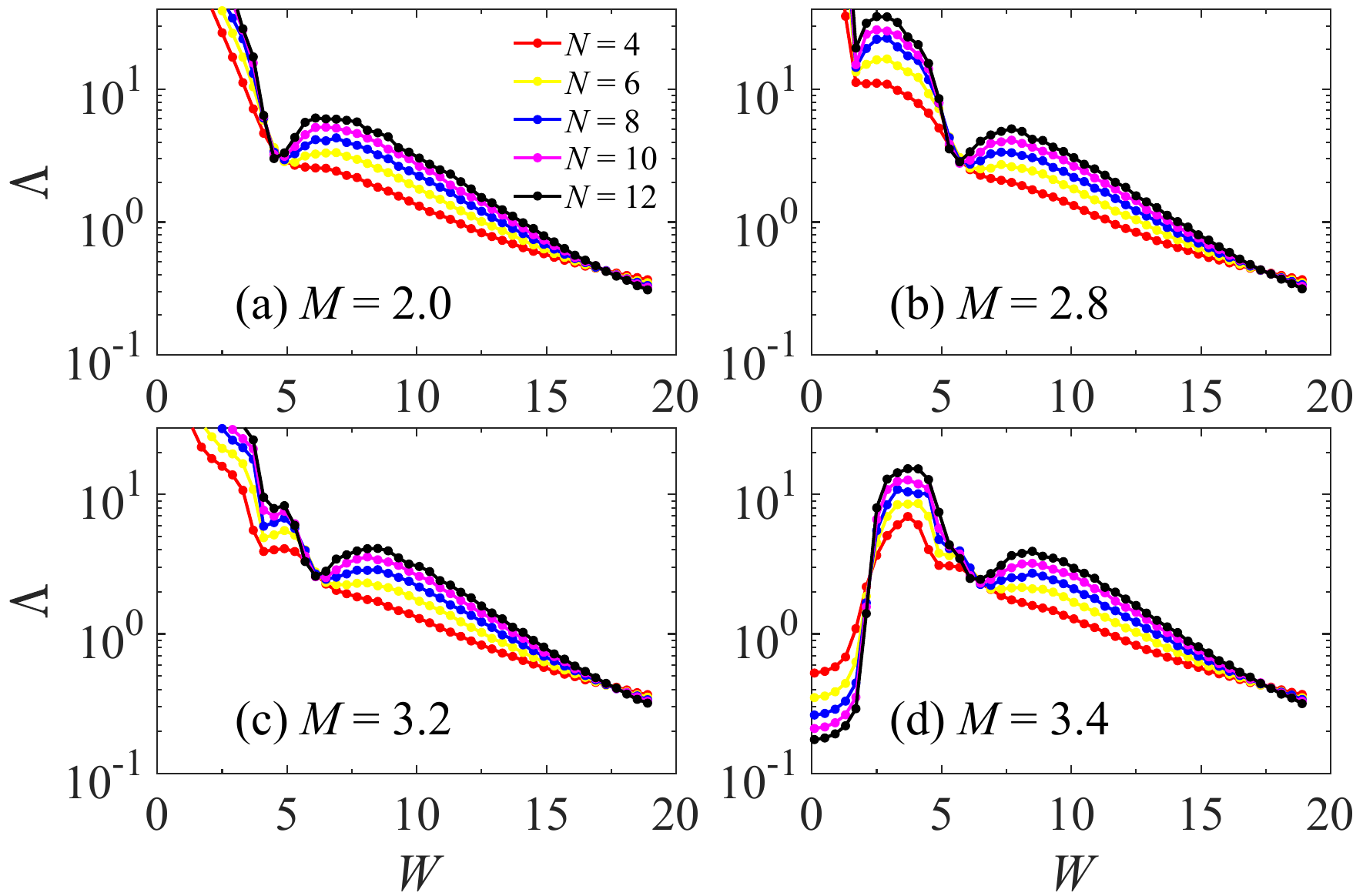}
	\caption{(Color online). Normalized localization length $\Lambda=\lambda(N)$ as a function of disorder strength $W$ at (a) $M=2.0$, (b) $M=2.8$, (c) $M=3.2$ and (d) $M=3.4$. Four cases are marked by the red circles in Fig.~\ref{fig5}. The curves correspond to different sample widths $N$.
		\label{fig6} }
\end{figure}

To study the disorder-driven phase transition, we calculate the localization length using the transfer matrix method \cite{MacKinnon1981,MacKinnon1983,Kramer1993,DYRPRB2023}. Considering a 3D finite $N\times N\times L$ sample with PBC in the $x$ and $y$ directions, the Lyapunov exponents (LEs), denoted as $\gamma_i$, reflecting the exponential decay behavior of the corresponding channels, can be calculated from the logarithms of the positive eigenvalues of the transfer matrix. The localization length $\lambda$ can be expressed with LE as $\lambda(N)=1/\gamma_{\mathrm{min}}(N)$
, where $\gamma_{\mathrm{min}}(N)$ is the smallest LE \cite{Beenakker1997,SJXiong2007} of the sample with width $N$. As $N$ increases, the normalized localization length $\Lambda\equiv\lambda(N)/N$ increases in a metallic phase, decreases in an insulating phase, and is unchanged at the phase transition point \cite{CCZPRL2015}.

Furthermore, the SCBA \cite{SJTPRB2012,CCZPRL2015,ZZQPRB2021A} is used to determine the band renormalization induced by weak and intermediate disorder. The renormalization of model parameters can be evaluated by self-energy
\begin{eqnarray}
\Sigma=\frac{W^2}{96\pi^3}\int\!\!\!\!\int\!\!\!\!\int_{\rm{BZ}}dk_xdk_ydk_z[\gamma_d(E_F+i0^+-H-\Sigma)^{-1}\gamma_d],\nonumber
\end{eqnarray}
where $\gamma_d=\rho_3\otimes\tau_0$ denotes the disorder preserves symmetries, and $E_F$ is the Fermi energy.
For simplicity, the self-energy can be decomposed as $\Sigma=\sum_{ij}\Sigma_{ij}\rho_i\otimes\tau_j$ with $i,j=0, 1, 2, 3$. Meanwhile, the Dirac mass $M$ is renormalized as $M_R=M+M_r=M+\Sigma_{01}$, where $M$ is bare mass without disorder, and the correction $M_r<0$ dominates the disorder-induced phase transition.

Our main results are summarized in the phase diagram on the $W-M$ plane in Fig.~\ref{fig5}.
There are five distinct phases
and the system with $M=3.4$ ($M=2.8$ or $3.2$) undergoes a sequence of NI-NLSM-HONLSM-DM-AI (NLSM-HONLSM-DM-AI) transitions with the increase of $W$.
The solid lines in Fig.~\ref{fig5} are determined by finite-size scaling of localization length,
while the dashed blue lines are the phase boundaries where the gap at zone center close [Figs.~\ref{fig2}(c) and \ref{fig2}(e)], predicted by SCBA.
The phase boundary determined by finite-size scaling of localization length [see red solid line in Fig.~\ref{fig5}] is consistent with the boundary of the NLSM and the NI obtained by SCBA [dashed  blue  line on the right side of Fig.~\ref{fig5}] under weak disorder. This strongly demonstrates the reliability of our results.

Figure~\ref{fig6} show the the scale-dependent behaviors of $\Lambda$ versus the disorder strength W for four typical $M$ [see the red circles in Fig.~\ref{fig5}].
In Fig.~\ref{fig6}(a), with $M=2.0$, the system initially resides in a metallic HONLSM phase characterized by $d\Lambda/dN>0$ under weak disorder.
Subsequently, it undergoes a transition to a diffusive metal (DM) phase, indicated by $d\Lambda/dN>0$, when $W$ exceeds a critical point at approximately $W\backsimeq4.6$, where $d\Lambda/dN=0$. Finally, after a metal-insulator transition occurs at $W\backsimeq17.2$, the system becomes localized due to strong disorder, transforming into an Anderson insulator (AI) \cite{AndersonPR1958}.
We note that the metal-insulator transition points  determine the solid lines in Fig.~\ref{fig5}.
By conducting this procedure at different values of $M$, one can determine the solid-line phase boundary shown in Fig.~\ref{fig5}.
Furthermore, the HONLSM-diffusive metal-AI multiple phase transition is also observed for
$M_z=2.8$ and $M_z=3.2$ when $W>5$. Meanwhile, a crossover from NLSM to HONLSM occurs, as indicated by a dip in the localization length within the region where $W<5$. This will be discussed later.
Additionally, we observe  a new type of multiple phase transitions in Fig.~\ref{fig6}(d): from a normal insulator to a NLSM, followed by a transition to a diffusive metal, and ultimately ending as an AI.



\begin{figure}[tbh]
	\centering
	\includegraphics[width=3.4in]{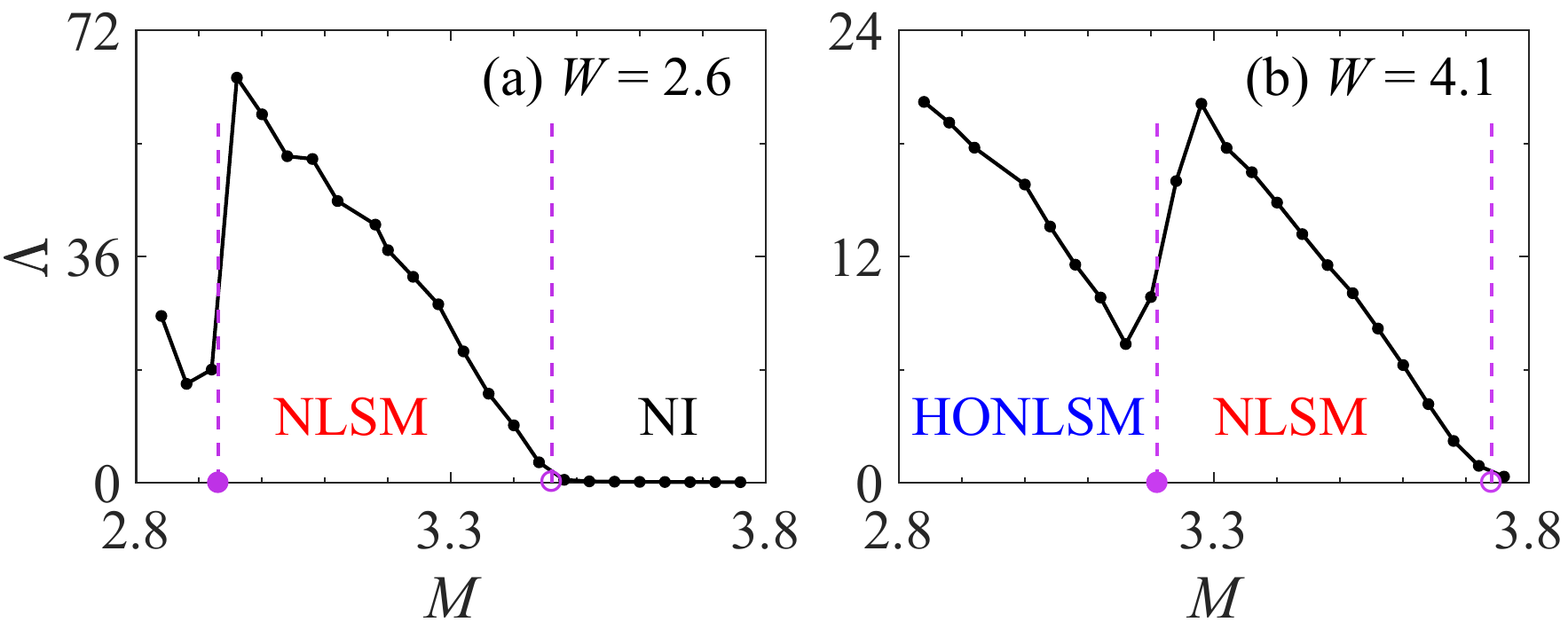}
	\caption{(Color online). Normalized localization length $\Lambda$ versus Dirac mass $M$ at (a) $W=2.6$ and (b) $W=4.1$, with sample widths $N=12$. The the violet dots and circles corresponding to those in Fig.~\ref{fig5}, among which the system is in NLSM state.
		\label{fig7} }
\end{figure}
Next, let us come to investigate the transition from a HONLSM to a NLSM.
As mentioned in Sec.~\ref{III}, the clean systems with $M=2.8$ and $3.2$ are in NLSM phase, and the transition to HONLSM phase at $M =3-\sqrt{2}\Delta$ is indicated by the split of one nodal ring in the momentum space. Nevertheless, there are no Anderson transition points indicated in the diagrams of $\Lambda$ [see Figs.~\ref{fig6}(b)  and \ref{fig6}(c)]. Here, the localization lengths are calculated with PBC in the $x$ and $y$ directions,  which means that the bulk transport properties of HONLSMs and NLSMs are analogous. 
Furthermore, we plot the normalized localization length as a function of $M$ for $W=2.6$ and $W=4.1$ in Fig.~\ref{fig7}.
As shown in Fig.~\ref{fig7}, there is a HONLSM-NLSM transition [violet dots]
around the dip of the renormalization localization length $\Lambda$.
The dip of $\Lambda$ found here is closely analogous to that in the transition from a higher-order Weyl semimetal to a Weyl semimetal \cite{ZZQPRB2021}.
However, the renormalization localization length at the dip is too large to be identified as an Anderson transition.
Continue to increase $M$, $\Lambda\rightarrow0$ at another transition point [see violet circles in Fig.~\ref{fig7}], representing the Anderson transition from a NLSM to a NI.


To verify our numerical results, in Fig.~\ref{fig5}, we illustrate the violet dots and circles that correspond to the transition points of the $\Lambda-M$ curves for various values of $W$ and compare them to the phase boundary determined by the SCBA theory.
As illustrated in Fig.~\ref{fig5}, under weak disorder, the transition points of $\Lambda-M$ curves (violet dots and circles) are consistent with the band-touching points obtained using SCBA (blue dashed lines), which further verifies the reliability of our numerical results. This also demonstrates that the HONLSM close to the NLSM  with nodal rings near the zone center is stable against weak disorder. This stability arises because disorder-induced renormalization of $M$ causes the nodal rings to repel each other, leading to their separation. This explains why the phase boundary is slanted toward the NLSM side in Fig.~\ref{fig5} and why there is an unexpected NLSM-HONLSM transition with increasing disorder. For a similar reason, the phase boundary of the NLSM and the NI is slanted toward the NI side.
\begin{figure}[tbh]
	\centering%
	\includegraphics[width=3.4in]{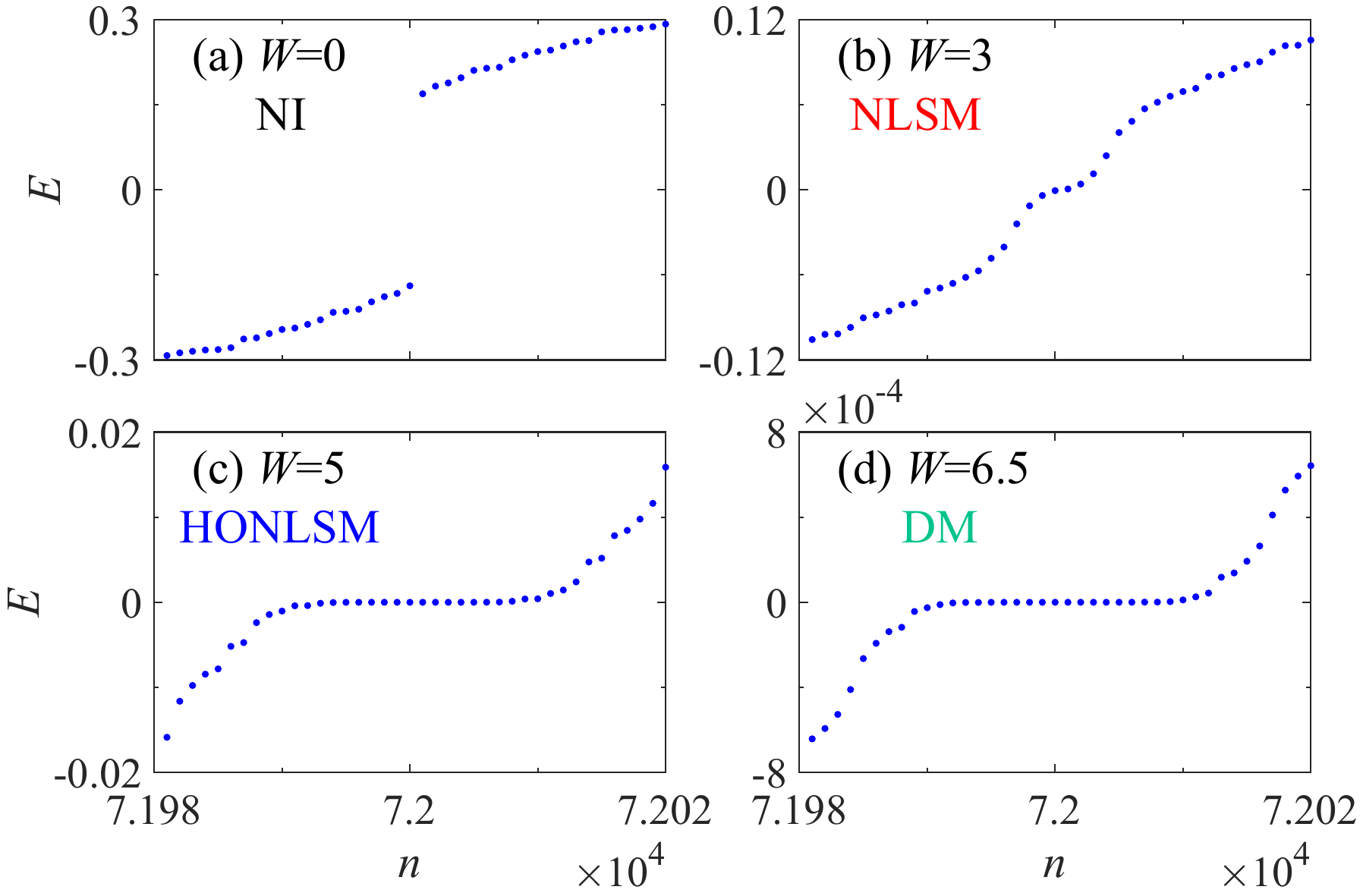}
	\caption{(Color online).  Eigenvalues of cubic samples at $M=3.4$ in (a) NI phase with $W=0$, (b) NLSM phase with $W=3$, (c) HONLSM with $W=5$, and (d) DM phase with $W=6.5$. The samples are under OBC in all three directions with sizes $N_x\times N_y\times N_z=60\times 60\times 10$, and only one disorder configuration is considered for each sample. 
		\label{fig8} }
\end{figure}

To further clarify the effects of disorder on band structure, we compare the disordered energy spectrum of the three metallic phases discussed above. In Fig.\ref{fig8}, we plot the eigenvalues of cubic samples under different disorder strengths $W$ when $M=3.4$. The clean system is in the NI phase and has a finite gap, as shown in Fig.\ref{fig8}(a). With the increase of $W$, some zero-energy states show up in Fig.\ref{fig8}(b) because the gap closes, and the sample transitions to an NLSM phase with one nodal ring. Continuing to increase $W$, the number of zero-energy states significantly increases, as illustrated in Fig.\ref{fig8}(c), since the system enters the HONLSM phase with hinge Fermi arc states with zero-energy flat bands. When $W=6.5$, the energy level spacing of the DM phase is much smaller than that of the HONLSM phase [see Figs.\ref{fig8}(c) and \ref{fig8}(d)]. Therefore, the evolution of the energy spectrum for $M=3.4$ is consistent with disorder-driven NI-NLSM-HONLSM-DM transitions in the phase diagram in Fig.\ref{fig5}.

\section{conclusion}
In summary, we have elucidated the topological characteristics of HONLSMs and clarified the phase transitions induced by disorder in HONLSM systems. Varying the Dirac mass parameter $M$ leads to distinct phases, including HONLSM, NLSM, and NI.
The hallmark of HONLSM phase is the existence of hinge Fermi arc states originate from the Fermi arcs that terminate on nodal rings, distinguishing it from the NLSM with only drumhead surface states.
In the presence of disorder preserves $PT$ and chiral symmetries, we obtain a global phase diagram utilizing the transfer matrix method and SCBA.
The Dirac mass decreases due to disorder renormalization, leading to the repulsion between nodal rings and causing disorder-driven transitions, such as NLSM-HONLSM-DM-AI transitions and NI-NLSM-HONLSM-DM-AI transitions.

\section*{Acknowledgments}
We thank Zhi-Qiang Zhang and Zi-Ming Wang for fruitful discussions. The authors acknowledge the support by the National Key R\&D Program of China (Grants No.
2022YFA1403700), the NSFC (under Grants No. 12074108, No. 11974256, and No. 12347101), the Natural Science Foundation of Chongqing (Grant No. CSTB2022NSCQMSX0568), the Natural Science Foundation of Jiangsu Province Grant (No. BK20230066), the Priority Academic Program Development (PAPD) of Jiangsu Higher Education Institution, Jiangsu Shuang Chuang Project (JSSCTD202209), and the Fundamental Research Funds for the Central Universities (Grant No. 2023CDJXY-048).

\appendix


\section{HONLSM phase for different $M$}\label{AB}
\begin{figure*}[tbh]
	\centering
	\includegraphics[width=6.8in]{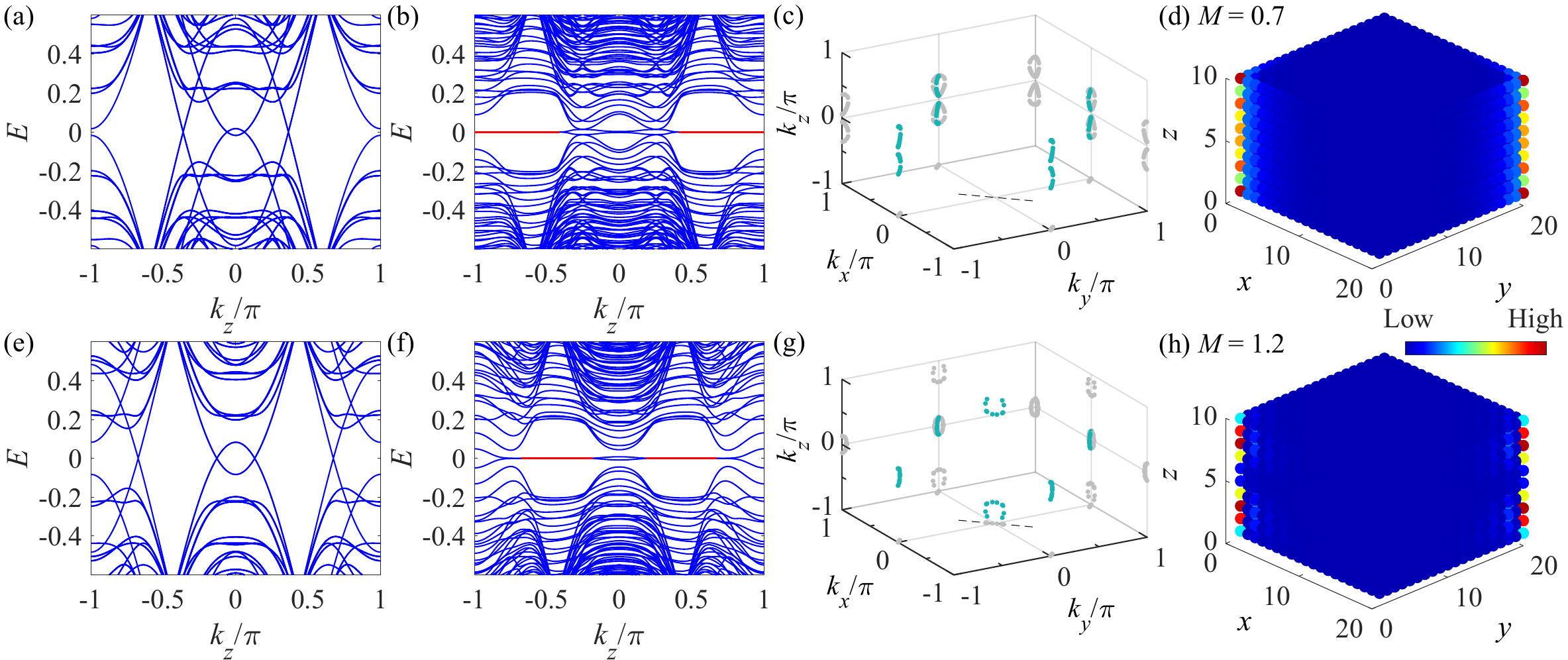}
	\caption{(Color online). Band structures and density of states DOS at $M=0.7$ [(a)-(d)] and $M=1.2$ [(e)-(h)]. The energy spectrum as a function of $k_z$ with periodic boundary conditions (PBC) for (a) and (e) and OBC for (b) and (f) in both $x$ and $y$ directions, respectively. $N_x=N_y=20$ in (a), (b), (e) and (f). The red lines in (b) and (f) represent the hinge Fermi arc states.
(c) and (g) Nodal rings in the 3D Brillouin zone. The gray rings are the projections of the nodal rings on the $k_x-k_y$, $k_y-k_z$, and $k_x-k_z$ planes. The dashed lines represent $k_y=-k_x$ direction. (d) and (h) DOS of the hinge Fermi arc states in HONLSM phase of a cubic sample with size $N_x\times N_y\times N_z=20\times 20\times 10$.
		\label{fig9} }
\end{figure*}
In the following, we show that the system is in the HONLSM phase when $0\leqslant M<3-\sqrt2\Delta$, where we set $\sqrt2\Delta<1$.

(i) As mentioned in Sec.~\ref{II}, $H_{\rm{mass}}$ will generate nodal rings around the Dirac nodes of the DSM with $\Delta=0$. When $1+\sqrt2\Delta<M<3-\sqrt2\Delta$, there are two complete nodal ring around the Dirac nodes located at $[0,0,\pm\arccos(M-2)]$. The rings connect into one nodal ring in the region $M\in(3-\sqrt2\Delta,3+\sqrt2\Delta)$, and the ring shrinks into a node at $M=3+\sqrt2\Delta$.

(ii) Considering that the gap at $\textbf{k}=(0, 0, \pi)$ in the energy spectrum of the Eq.~(\ref{eq2}) is closed at $M=1\pm\sqrt2\Delta$, we discuss the $0\leqslant M<1-\sqrt2\Delta$ regime and the $M\in(1-\sqrt2\Delta,1+\sqrt2\Delta)$ regime, separately.
For $0\leqslant M<1-\sqrt2\Delta$,  the nodal rings are around the nodes at $[0,\pi,\pm\arccos M]$ and $[\pi,0,\pm\arccos M]$ [see Figs.~\ref{fig9}(c) and \ref{fig1}(b)-\ref{fig1}(c)].
On the other side, for $M\in(1-\sqrt2\Delta,1+\sqrt2\Delta)$
as depicted in Fig.~\ref{fig9}(g), there are nodal rings exist around the Dirac nodes lie $[0,\pi,\pm\arccos M]$, $[\pi,0,\pm\arccos M]$ and $[0,0,\pm\arccos(M-2)]$ simultaneously [see also Figs.~\ref{fig1}(b) and \ref{fig1}(c)].
Nevertheless, even if the positions of the nodal rings in the region $M\in[0,1+\sqrt2\Delta)$ differ from those in the main text, hinge Fermi arcs persist between nodal rings under OBC in the $x$ and $y$ directions [see Figs.~\ref{fig9}(a), \ref{fig9}(b), \ref{fig9}(e), and \ref{fig9}(f)]. The presence of hinge Fermi arc states is observable in a cubic sample [see Figs.~\ref{fig9}(d) and \ref{fig9}(h)].

\bibliographystyle{apsrev4-2} 
\bibliography{ref_HONLSM}

\end{document}